\documentclass[aps,showpacs,amsmath,amssymb,prl,groupedaddress, floatfix,twocolumn]{revtex4-1}

\usepackage{graphicx}
\usepackage{dcolumn}
\usepackage{bm}
\usepackage{amssymb}
\usepackage{multirow}
\usepackage{SIunits}
\usepackage{float}
\usepackage{todonotes}
\usepackage{color}  
\usepackage{todonotes}

\newcommand{\mt}[1]{\mathrm{#1}} 
\newcommand   {\<}{ \langle }
\renewcommand {\>}{ \rangle }

\newcommand{\ket}[1]{|#1\rangle}

\begin{document}

\title{Unraveling the Structure of Ultracold Mesoscopic Molecular Ions}

\author{J. M. Schurer}
    \email{jschurer@physnet.uni-hamburg.de}
\author{A. Negretti}
\author{P. Schmelcher}
    \email{pschmelc@physnet.uni-hamburg.de}

\affiliation{Zentrum f\"ur Optische Quantentechnologien, Universit\"at Hamburg, Luruper Chaussee 149,
22761 Hamburg, Germany}
\affiliation{The Hamburg Centre for Ultrafast Imaging, Universit\"at Hamburg, Luruper Chaussee 149, 22761
Hamburg, Germany}

\date{\today}


\begin{abstract}

We present an in-depth many-body investigation of the so-called mesoscopic molecular ions 
that can build-up when an ion is immersed into an atomic Bose-Einstein condensate in one dimension. To 
this end, we employ the Multi-Layer Multi-Configuration Time-Dependent Hartree method for Mixtures of 
ultracold bosonic species  for solving the underlying many-body Schr\"odinger equation. This enables us to 
unravel the actual structure of such massive charged molecules from a microscopic perspective. 
Laying out their phase diagram with respect to atom number and interatomic interaction strength, we 
determine the maximal number of atoms bound to the ion and reveal spatial densities 
and molecular properties. Interestingly, we observe a strong interaction-induced localization, 
especially for the ion, that we explain by the generation of a large effective mass, similarly to ions in 
liquid Helium. Finally, we predict the dynamical response of the ion to small perturbations.
Our results provide clear evidence for the importance of quantum correlations, as we demonstrate by 
benchmarking them with  wave function ansatz classes employed in the literature.

\end{abstract}

\maketitle

\paragraph*{Introduction.--}

In early studies on ions in liquid $^4$He, a small ionic mobility in the liquid was detected 
experimentally~\cite{Meyer1958}. To explain this observation, a high liquid density around the ion was 
suggested~\cite{Atkins1959}. The latter was subsequently corroborated by the generation of a large 
effective mass for the ionic impurity~\cite{Gross1962}, as many atoms are attracted to the ion. In recent 
years, the combination of degenerate quantum gases and trapped ions has opened new 
perspectives~\cite{Harter2014} thereby allowing to explore the 
underlying mechanisms of such phenomena. Indeed, the exquisite controllability of both quantum gases and 
trapped ions enable in-depth investigations of fundamental processes ranging from ultracold chemical 
reactions~\cite{Ratschbacher2012,Rakshit2011,Harter2012a}, charge transport~\cite{Cote2000a}, and spin 
decoherence~\cite{Ratschbacher2013} to sympathetic cooling~\cite{Zipkes2011a,Harter2012a,Meir2016a} and the 
strong-coupling regime of polaron physics~\cite{Casteels2010}.
Importantly, the atom-ion interaction supports the formation of weakly-bound charged dimers with binding 
radii of hundreds of nanometers or more~\cite{Cote2002} which can be formed by three-body 
collisions~\cite{Krukow2016,Krukow2016a} or radiative processes~\cite{Rakshit2011}. 
These molecules are reminiscent of Feshbach or halo molecules~\cite{Krems2009,Quemener2012,Stipanovic2014}
as their neutral counterparts are named and represent an example for extraordinary  molecules
with a binding radius and a de-Broglie wave length of the same order of magnitude.  
Even more fascinating, they can consist of multiple bosonic atoms and a single ion, such that they 
become mesoscopic massive quantum objects~\cite{Cote2002}, eventually even exhibiting a shell 
structure~\cite{Gao2010}. 
When no population of the bound states occurs, a single tightly confined ion is predicted to 
induce a micron sized density disturbance with hundreds of excess atoms in an ultracold 
gas~\cite{Gross1962,Massignan2005} which becomes a clear density hole 
in the Tonks-Giradeau limit~\cite{Goold2010}. However, such a density hole increasingly
closes if bound states become populated~\cite{Schurer2014}.

\begin{figure}[t]
 \centering
 \includegraphics[width=1.0\linewidth]{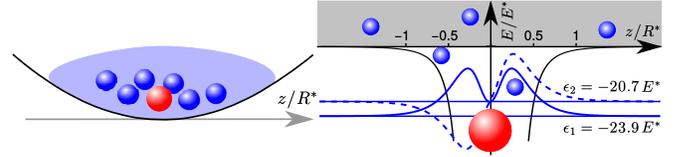}
 \caption{(Color online) Setup and atom-ion interaction. (Left) Atoms (blue) and ion (red) in a 
quasi one-dimensional harmonic trap. (Right) Atom-ion interaction potential (solid black line) together with 
the two most weakly bound states and their energies $\epsilon_i$; bound and unbound atoms are indicated.} 
\label{fig:AIinter}
\end{figure}

In this work, we explore the quantum state of such mesoscopic molecular ions in one spatial dimension (1D) 
based on a microscopic theory (see Fig.~\ref{fig:AIinter}). Thereby, we are able to derive a complete 
zero-temperature phase-diagram for the compound system and show how strongly the critical cluster 
size~\cite{Cote2002} is affected when particle correlations are taken into account. We confirm the 
hypothesis of interatomic interaction-induced excitations stabilizing the molecular 
cluster~\cite{Gao2010} and observe as well as explain the self-localization behavior of the ion, which 
becomes possible by incorporating the ionic motional degree of freedom, not taken into account in earlier 
studies~\cite{Massignan2005,Goold2010,Gerritsma2012,Schurer2014,Schurer2015,Schurer2016}. As a result, 
we connect the atomic density profiles, in particular the predicted density hole~\cite{Massignan2005, 
Goold2010,Schurer2014}, with the spatial extent of the ionic wave function and predict the dynamical response 
of the ion to a small perturbation. All this is attained by exploiting the knowledge of the numerically 
computed many-body correlated quantum state of the compound system and allows to benchmark commonly used wave 
function ansatz classes.


\paragraph*{Setup.--}
We study a single ion of mass $m$ and position $z_\mt{I}$ immersed into a cloud of $N$ bosonic 
ultracold atoms of mass $m$ located at $z_i$ both confined in a harmonic trap of frequency $\omega$. 
Let us remark that the choice of equal trap frequencies is for reasons of simplicity and our results (see 
below) do generalize to the case of unequal trapping frequencies. The atoms interact via a 
contact-interaction potential of strength $g$, while the atom-ion interaction is given by 
$V_\mt{AI}( z_i -  z_\mt{I}) = -\frac{1}{2}\alpha e^2( z_i -  z_\mt{I})^{-4}$~\footnote{
See Supplemental Material at [URL] for details on the atom-ion interaction, the frame transformations, the 
employed models, the effective force analysis, and for a discussion of convergence of our computational 
approach.} 
with the atomic polarizability $\alpha$ inducing a characteristic 
length $R^* =\sqrt{\alpha e^2 m/(2\hbar^2)} $ and energy $E^* = \hbar^2/(m {R^*}^2)$ scale. Moreover, we 
take the two most weakly bound states of the atom-ion interaction into account (see Fig.~\ref{fig:AIinter}) 
which are eigenstates of the relative Hamiltonian $-(\hbar^2/m)\partial_r^2 + V_\mt{AI}(r)$ 
with energy $\epsilon_i$. In order to reveal the physics originating from the atom-ion interaction, we set the 
trap length $l = \sqrt{\hbar/(m\omega)} = R^*$.

\begin{figure}[t]
 \centering
\includegraphics[width=1.0\linewidth]{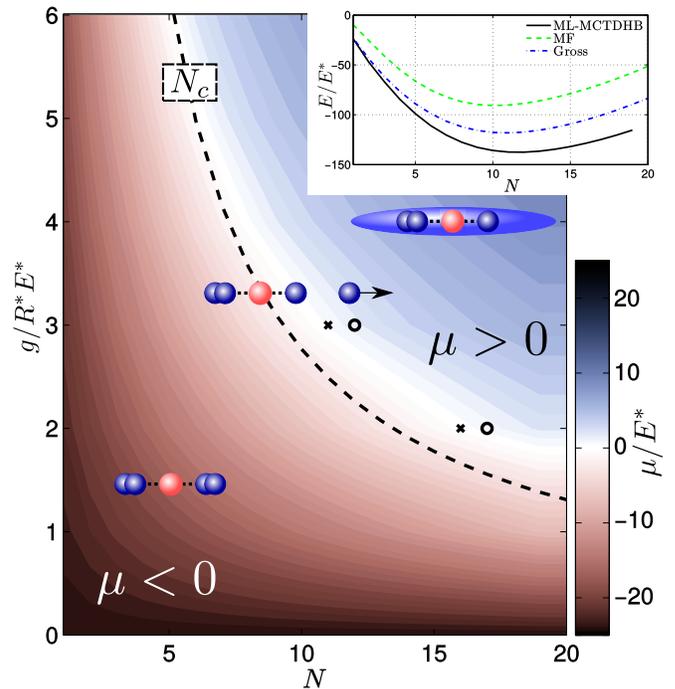}
 \caption{(Color online) Phase diagram. Chemical potential $\mu$ from the Gross ansatz 
as a function of $N$ and $g$. Black circle (crosses) 
mark $\mu=0$ from ML-MCTDHB (Gross). The black dashed line presents the estimation $g_c \approx (\omega - 
\epsilon_1) /(N_c-1)$. 
(Inset) Total energy $E(N)$ as a function of $N$ for $g=3E^*R^*$.
} \label{fig:phaseDiag}
\end{figure}

\paragraph*{Phase Diagram.--}
Depending on $N$ and $g$, two distinct phases for the ground state occur (see 
Fig.~\ref{fig:phaseDiag}), which can be separated by looking at the sign of the chemical potential 
\begin{equation}
 \mu = E(N+1,g) - E(N,g) 
\end{equation}
with the total energy $E(N,g)$. For $\mu<0$, 
the presence of the bound states makes the binding of all bosonic atoms possible such 
that a single \emph{mesoscopic charged molecule} is formed. The near linear decrease of $E(N)$ (inset) 
shows that the atoms are ``inserted one by one'' into the bound state, which is only possible due to their 
bosonic nature.
In contrast, for $\mu>0$, the total energy can not be reduced anymore by adding another atom. This clearly 
indicates that not all atoms can be bound, since the ion becomes increasingly screened, resulting in an 
unbound, yet trapped, atomic fraction. In between 
these two regimes, the \emph{dissociation} of the molecule occurs at $\mu = 0$ defining the maximal number of 
atoms $N_c$ that can be bound to the ion for a fixed $g$. Hence, we find a transition from an all-bound 
many-body state to a molecule immersed into an unbound background gas. One can estimate the threshold region 
by energetic considerations to be $g_c \approx (\omega - \epsilon_1) /(N_c-1)$ (see Fig.~\ref{fig:phaseDiag} 
dashed line).

The question which arises now is: How to capture the essential nature of such a many-body quantum state, 
particularly from microscopic considerations ? A natural starting point for the theoretical description of 
the wave function $\ket{\Psi}$ is obtained by variationally optimizing a product 
ansatz
\begin{align}\label{eq:prodState}
 \Psi_\mt{MF}(z_\mt{I},z_1,\cdots,z_N) &= \varphi(z_\mt{I})\prod_{i=1}^N \chi(z_i), \quad \text{or}\\
 \Psi_\mt{G}(Z_\mt{I},Z_1,\cdots,Z_N) &= \varphi(Z_\mt{I})\prod_{i=1}^N \chi(Z_i).
\end{align}
The first ansatz $\Psi_\mt{MF}$ corresponds to a product of the atomic and the ionic part of the wave 
function together with a Gross-Pitaevskii ansatz for the atomic part. Hence, we refer to this ansatz as 
mean-field (MF). The second ansatz $\Psi_\mt{G}$, inspired by Gross~\cite{Gross1962}, is a product in the 
ion-frame (IF) coordinates $Z_\mt{I} = z_\mt{I}$ and $Z_i = z_i - z_\mt{I}$. 
In order to go even beyond both ansatz wave functions, we employ the multi-layer 
multi-configuration time-dependent Hartree method for bosons (ML-MCTDHB) \cite{Kronke2013,Cao2013} (see 
Supplemental Material~\cite{Note1}), which allows us to numerically compute the ground state of the hybrid 
system via imaginary time-propagation, i.e.\ relaxation.
We observe that the MF can reproduce a minimum in the total 
energy (see inset of Fig.~\ref{fig:phaseDiag}), nevertheless it predicts a substantially too large energy. 
The Gross ansatz already lowers the total energy hence is closer to the true ground state due to the 
underlying  variational principle. The ML-MCTDHB results, however, further approach the true many-body 
ground state such that we can use them to benchmark the MF and the Gross approach.
In addition to the lowering of $E(N)$, it also predicts the dissociation at larger $N$ (c.f.\ circle and 
crosses).

\begin{figure}[t]
 \includegraphics[width=1.0\linewidth]{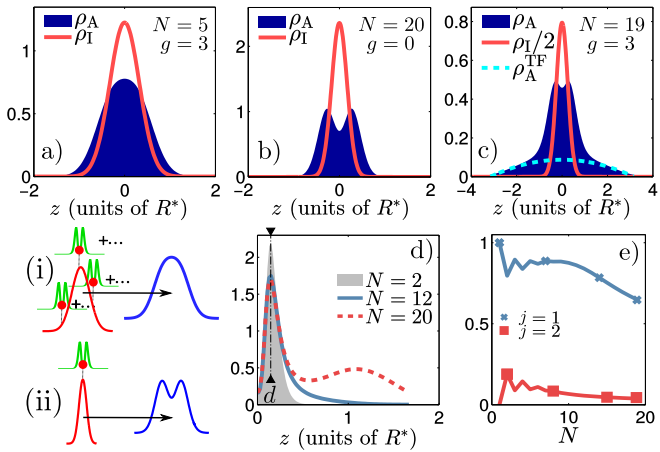}
 \caption{(Color online) Molecular structure. (a-c) Atomic (shaded) and ionic (solid line) density 
profiles. In (c) also a Thomas-Fermi profile with $N-N_c$ particles is shown (dashed 
line). (d) Atom-ion correlation function $g_2(z)$ for $g=3R^*E^*$. (e) Population of the bound states 
$f_j/N$. (i) Delocalized versus (ii) localized ion and its impact on the atomic density.
} \label{fig:structure}
\end{figure}

\paragraph*{Molecular Structure.--}
In order to unravel the structure of such a many-body state, we begin with the atomic and ionic 
density profiles $\rho_\mt{I(A)}(z) = \< \hat \Psi_\mt{I(A)}^\dagger(z) \hat \Psi_\mt{I(A)}(z)\>$ (see 
Fig.~\ref{fig:structure}) where $ \hat \Psi_\mt{I(A)}$ are the ion (atom) field operators. We observe that 
for small $N$ both density distributions are of similar shape and spatial extension though with different 
maximal values. For large $N$, the ion 
becomes significantly localized, while the atoms reveal two peaks in the density ($g=0$) 
and exhibit the formation of broad shoulders ($g>0$). We observe that
the qualitative behavior of the latter can be captured by a Thomas-Fermi (TF) profile with $N_\mt{TF} = 
N-N_c$ atoms (cyan line). However, we emphasize that the atoms are strongly correlated and far 
away from the validity regime of the TF approximation.
The fact that the atoms are bound or unbound is, however, not obvious from the density profiles and 
becomes only explicit in the atom-ion correlation function 
\begin{equation}
 g_2(z) = \frac{\< \hat \Psi_\mt{I}^\dagger(z) \hat \Psi_\mt{A}^\dagger(-z) \hat 
\Psi_\mt{A}(-z) \hat \Psi_\mt{I}(z)\>} {N \rho_\mt{I}(z) \rho_\mt{A}(-z)}
\end{equation}
with $z = z_\mt{A}-z_\mt{I}$ shown in Fig.~\ref{fig:structure} d). Here, we can clearly see that it is most 
likely to find an atom at the binding distance $d$ (vertical dashed line) from the ion, while larger 
distances are strongly suppressed for $N<N_c$ ($N_c=12$ for $g=3E^*R^*$). Note that the MF ansatz results in 
$g_2(z_\mt{A}-z_\mt{I}) = 1$, i.e.\ no binding is possible. 
The atomic density profile can now be explained by sampling the $g_2(z)$ profile with well-defined 
binding distance over the ionic density distribution (see sketch in Fig.~\ref{fig:structure}). While for (i) 
a spatially spread ion the molecular structure is hidden by the sampling, (ii) a 
localized ion reveals details of the binding by the two density peaks representing the strong 
bunching at distance $d$. In this way, we rediscover the onset of the central density hole predicted for a 
static ion~\cite{Goold2010,Schurer2014}, however, here it is induced by the atom-ion interaction instead of 
originating from an external strong confinement.
For $N$ approaching $N_c$, one  observes that the atom-ion correlation function broadens to larger relative 
distances which reduces the bunching at $d$. This corresponds to a spatial increase of the 
bound-state width. Beyond the dissociation point $N_c$, the strong suppression of larger atom-ion distances 
is lifted and the occurrence of the unbound fraction becomes prominent [see second maximum in 
Fig.~\ref{fig:structure} d)]. 

While the Gross ansatz is able to reproduce this behavior of $g_2(z)$ qualitatively, it does not allow for 
population of an odd state due to the parity symmetry of the ground state. In Fig.~\ref{fig:structure}, the 
population of the two bound states $f_j = \< \hat a^\dagger_{j} \hat a_{j}\>$ is shown with $\hat a_{j}$ 
($\hat a^\dagger_{j}$) being their annihilation (creation) operators. We find a significant population of the 
second bound state in particular for even $N$. This excitation of atoms to the more weakly bound state allows 
to reduce the inter-atomic repulsive energy and hence stabilizes the many-body bound state. This explains the 
observed increase of $N_c$ obtained from 
the correlated ML-MCTDHB results and can be viewed as the 1D analog of shell structure formation. 

\begin{figure}[b]
\includegraphics[width=1.0\linewidth]{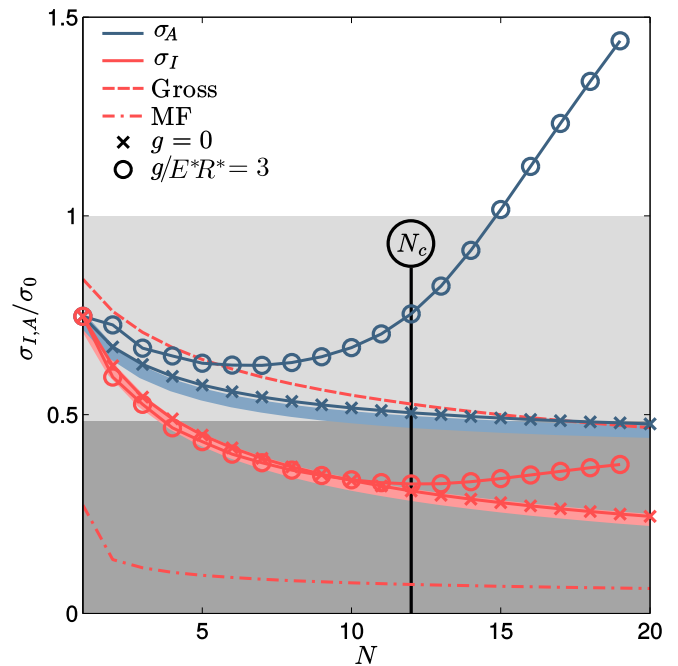}
 \caption{(Color online) Self-localization. 
Variance of the atomic (blue) and the ionic (red) variance normalized by the non-interacting 
variance $\sigma_0$. 
Ionic variance from the Gross (dashed) and  the mean-field (dashed dotted) ansatz for $g=0$ are shown, too. 
The dark (light) thick lines represent $\sigma_\mt{I}$ ($\sigma_\mt{A}$) solely including 
the increase of $M$~\cite{Note1}. Dark (light) gray area represents the 
spatial extent of the bound states ($\sigma_0$). 
} \label{fig:rms}
\end{figure}

\paragraph*{Self-Localization.--}
As previously seen, the increase of $N$ localizes the ion. For 
a more quantitative analysis, we use the atomic ($\sigma_\mt{A}^2 = \<\frac{1}{N}\sum_{i=1}^N 
z_i^2\>$) and the ionic ($\sigma_\mt{I}^2 = \<z_\mt{I}^2\>$) variance, shown in Fig.~\ref{fig:rms} in units 
of the non-interacting variance $\sigma_0 = l/\sqrt{2}$. 
Already for $N<N_c$, we observe that the ion as well as the atoms localize on a length 
scale smaller than the trap length. Since this is solely induced by the atom-ion interaction, we call 
it self-localization.
For $g=0$, both variances decrease monotonously with increasing $N$. Since in this case the state is an 
$(N+1)$-body cluster, we can understand this self-localization solely by the increase of the total mass 
$M=(N+1)m$ localizing the center of mass wave function of the complete atom-ion system (thick lines; see 
Supplemental Material~\cite{Note1}). In this 
way, the atomic variance approaches the width of the bound state (dark gray area) because the static ion
assumption becomes increasingly valid. Be aware that while the MF (dashed dotted line) strongly underestimates 
the ionic variance, the Gross ansatz strongly overestimates it (dashed line).
For $g>0$, the variance $\sigma_\mt{A}$ reveals a minimum and increases already for $N < N_c$. This goes 
hand in hand with the spatial widening of $g_2$ [solid line in Fig.~\ref{fig:structure} d)] which 
we interpret as a broadening of the 
bound state. We emphasize that only when the effective bound state variance becomes comparable to the 
trap length the impact of the confinement on the molecular ion goes beyond the localization of the center of 
mass.  
In this case, one might think of a ``molecule under pressure''~\cite{Baye2008,Laughlin2009}.
Beyond $N_c$, the ionic self-localization is reduced while the emergence of the shoulders in the 
atomic density gives rise to a rapid increase of $\sigma_\mt{A}$. 
Here it becomes evident that the situation of equal trapping frequencies for atoms and ion 
represents no restriction to the generality of our results. Since the atoms and in particular the ion 
localize on distances smaller than their trapping length, the confining potentials only determine the center 
of mass variance. However, 
the atomic trap becomes indeed important for $N>N_c$, impacting the dissociation and defining the spatial 
extent of the unbound fraction. In contrast, the ion trap has actually vanishing impact such that it could in 
most cases even be switched off. 


\paragraph*{Low energy excitations.--}
In order to learn about the dynamical response of the strongly correlated ion 
within the bosonic ensemble to e.g.\ a quench of the ionic trap frequency, we introduce 
an effective single particle of mass $m^*$ confined in a harmonic trap
of frequency $\omega^*$~\footnote{As shown in Ref.~\cite{Catani2012} a trapped impurity 
may experience an effective potential.}. Motivated by an ion density profile which is very well 
approximated by a Gaussian~\cite{Note1} and minor correlations between the ionic and the atomic 
IF coordinates, we use the particle associated to the ionic variable $Z_\mt{I}$ in the IF as
effective particle. By construction, it has the equivalent density profile as the ion itself such
that spatial measurements can be associated to both of them. From the Gross ansatz, one expects
a free particle of mass $m$ in a trap $\omega\sqrt{1+N}$.
In order obtain the effective frequency $\omega^*$, one could excite a breathing 
oscillation~\cite{Catani2012}.
Here, however, we compute $\omega^*$ from the spatial width $l^* = \sqrt{\hbar/(m^*\omega^*)} 
(= \sqrt{2}\sigma_\mt{I})$ and the effective force exerted on the effective particle. Employing the 
knowledge of the full many-body wave function, the effective force 
$F_\mt{I}^*(Z_\mt{I})$ is given by the partial trace of the force operator $F_\mt{I} = -[\partial_{Z_\mt{I}}, 
H]$ with $H$ being the total system Hamiltonian in the IF~\cite{Note1}.

\begin{figure}
\includegraphics[width=1.0\linewidth]{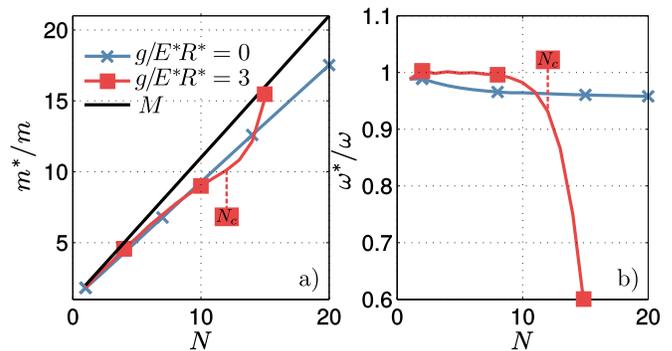}
 \caption{(Color online) Effective ion behavior. a) Effective mass $m^*/m$  and b) effective trapping 
frequency $\omega^*/\omega$. Note that the Gross ansatz gives $\omega_\mt{G} = 
\omega\sqrt{1+N}$  and $m_\mt{G} = m$.
} \label{fig:effParticle}
\end{figure}

The resulting $m^*$ and $\omega^*$ are shown in Fig.~\ref{fig:effParticle}.
In case all atoms are bound ($N<N_c$), we observe that the effective ion accumulates a large 
mass, nearly the total mass $M$, increasing linear in $N$, while $\omega^*$ is 
varying very little and is given approximatively by the trap frequency. Hence, the localization can be 
understood 
by the generation of a huge effective mass. Approaching $N_c$ for $g/E^*R^*=3$, $m^*$ becomes sub-linear 
whereas for $N>N_c$ it rapidly increases even to the total mass $M$. At the same time, the 
effective frequency strongly decreases revealing a slow response. Note that we do not
give $\omega^*$ and $m^*$ for $g=3$ beyond $N=15$ because here the effective single particle 
picture breaks down~\cite{Note1}. We remark that small effective trapping frequencies and large effective 
masses are reminiscent of the behavior found for the ionic polaron in the strong coupling 
regime~\cite{Casteels2010}.

\paragraph*{Discussion and Experimental Realization.--}
The attainment of the ultracold $s$-wave collision regime in atom-ion systems is under intense 
investigation~\cite{Ciampini2002,Grier2009,Schmid2010,Zipkes2010,Ravi2012a,Lambrecht2016}. The hybrid system 
can be created either by combination of atom and ion traps~\cite{Grier2009} or by fast ionization 
of a few atoms~\cite{Ciampini2002}.
For instance, assuming a $^{87}$Rb$^+$ ion in a $^{87}$Rb atomic cloud ($R^* = 
260\nano\meter$ and $E^*/h=1.6\kilo\hertz$), our setup corresponds to a trap frequency
of $\omega \approx 2\pi\cdot1.6\kilo\hertz$. With a transversal trapping frequency of $\omega_\bot \approx 
2\pi\cdot50\kilo\hertz$, we obtain $g\approx 1 E^*R^*$~\cite{Olshanii1998}. 
The formation of molecular ions, however, now relies on the occurrence of three-body collisions, the dominant 
reaction channel already at moderate densities~\cite{Krukow2016a}, or can be induced by 
either photo-association~\cite{Rakshit2011} or a Raman-type scheme~\cite{Cote2002}.
In this work, we have assumed that only the two most weakly bound states are of relevance for this 
reaction. This can be justified by the strong suppression of direct atom-capture into more 
deeply bound states~\cite{Cote2002}. Even though, these processes dictate the life-time of the molecular 
ion. 
Once the molecule is formed, it can be probed by measuring the atomic excess density near 
the ion~\cite{Massignan2005,Goold2010} or by wave-guide expansion~\cite{Schurer2014}.
Moreover, the binding can be identified via the effective ion mass by measuring the ionic 
variance~\cite{Gerritsma2010} in the ground-state and during breathing dynamics.


\paragraph*{Conclusions.--}
We have derived and characterized the many-body bound state of $N$ atoms and a single ion both confined 
in a harmonic trap. The dissociation threshold $N_c$ has been identified, for which a transition from an 
all-bound molecular ion to a molecule immersed into a background gas takes place. We have seen that even 
though the spatial extend of the particles is larger than the binding distance, one can identify the binding. 
The latter induces a substantial self-localization behavior for atoms and ion. 
Beyond that, we showed that the ion behaves like an effective particle of nearly the total 
mass in the bare ion trap.
In addition, we were able to benchmark simplistic wave function classes via the ML-MCTDHB method, 
showing that correlations counteract the localization and stabilize the molecular cluster.
Our results can be viewed as the basis for future intriguing studies concerning mesoscopic many-body
bound states. A promising direction concerns the molecular formation process which can give a handle on 
formation time scales and stability. Directly related is the question regarding the energy and the structure 
of internal molecular excitations. Moreover, the insights gained into the structure of the 
many-body wave function can stimulate the design of a unifying, simple and predictive, theoretical model 
which captures the essential physics over the complete parameter regime even up to high atom numbers.


\begin{acknowledgments}
\paragraph*{Acknowledgements.--}
The authors acknowledge inspiring conversations with Igor Lesanovsky and  Zbigniew Idziaszek and helpful 
advice by Rene Gerritsma. Moreover, the authors thank Juliette Simonet for a detailed feedback on the 
manuscript. J.S. also thanks Sven Kr\"onke and Valentin Bolsinger for many clarifying 
discussions and valuable suggestions.
This work has been financially supported by the excellence cluster 'The Hamburg Centre for
Ultrafast Imaging - Structure, Dynamics and Control of Matter at the Atomic Scale' of
the Deutsche Forschungsgemeinschaft.
\end{acknowledgments}



\bibliography{library}


\end{document}